\documentclass[preprint,12pt]{elsarticle}

\usepackage{xcolor}

\usepackage{mathrsfs}
\usepackage{amsmath}

\usepackage{amsthm}
\usepackage{amssymb}

\usepackage{listings}
\usepackage{color} 
\definecolor{mygreen}{RGB}{28,172,0} 
\definecolor{mylilas}{RGB}{170,55,241}

\journal{Physica A}

\begin{document}

\begin{frontmatter}

\title{When is sync globally stable in  sparse networks of \\ identical Kuramoto oscillators? }

\author[Bard]{Yury Sokolov\corref{cor1}\fnref{newAddress}}%
\ead{ysokolov@ucsd.edu}
\author[Bard]{G. Bard Ermentrout}
\ead{bard@pitt.edu}

\cortext[cor1]{Corresponding author}

\address[Bard]{Department of Mathematics, University of Pittsburgh, Pittsburgh, PA 15260, USA}

\fntext[newAddress]{Present address: Department of Medicine, UC San Diego, La Jolla, CA 92037, USA}

\date{\today}

\begin{abstract}
Synchronization in systems of coupled Kuramoto oscillators may depend on their natural frequencies, coupling, and underlying networks. In this paper, we reduce the alternatives to only one by considering identical oscillators where the only parameter that is allowed to change is the underlying network. While such a model was analyzed over the past few decades by studying the size of the basin of attraction of the synchronized state on restricted families of graphs, here we address a qualitative question on general graphs. In an analogy to resistive networks with current sources, we describe an algorithm that produces initial conditions that are often outside of the basin of attraction of the synchronized state. In particular, if a graph allows a cyclic graph clustering with a sufficient number of clusters or contains a sufficiently long induced subpath without cut vertices of the graph then there is a non-synchronous stable phase-locked solution. Thus, we provide a partial answer to when the synchronized state is not globally stable.  

\end{abstract}




\end{frontmatter}


\section{Introduction}
The question of synchronization in networks of coupled oscillators has played an important role in the development of nonlinear science with wide range of applications and areas that stimulated productive work during the last decades. This ranges from physics, engineering, mathematics to biology, where synchronization appears on different levels of organization from flocks of birds to neuronal dynamics. 

One of the canonical models in the study of synchronization is the Kuramoto model describing the dynamics of a network of coupled oscillators using a simple phase model. While first attempts in the analysis of synchronization phenomena required strong assumptions about the system, e.g., the underlying network of coupled oscillators has to be a complete graph \citep{KuramotoPaper}, the developments in network science and nonlinear dynamics over the past decades have allowed us to gain a deeper insight into this problem. 

It has been observed that for some graphs in addition to synchronized state (sync) the Kuramoto model may have other stable equilibria such as twisted states, which occupy the phase space, and as a consequence, reduce significantly (positive measure) the the basin of attraction of sync \citep{Wiley}. 

During the last decades, the major focus has been on the (local/linear) stability of sync and twisted states. In particular, conditions for the existence and uniqueness of a stable equilibrium with phase differences at most $\pi/2$ in the Kuramoto model with arbitrary natural frequencies on a graph has been shown in \citep{DorflerBullo}. A sufficient condition for the instability of an equilibrium in terms of graph cuts has been derived in \citep{Mallada}. In \citep{Bronski}, authors have showed the connection between the instability of equilibria and the cycle space of the underlying graph. The last study has been extended in \citep{Ferguson}, where the author illustrates, in particular, how equilibria correspond to the lattice points in a certain set. The stability of twisted states of the Kuramoto model defined on Cayley and Erd\H{o}s-R\'enyi random graphs is studied in \citep{Medvedev}. 

In addition, several attempts have been made to estimate how big the basin of attraction of sync is \citep{Wiley, Jacquod17}. In contrast, in this paper we provide a partial answer to a qualitative question, i.e., whether sync is globally stable or not. We do not address here a quantitative question: what is the size of the sync basin. 


%
%

We consider a simplified Kuramoto model on a sparse graph $G$ of order $n$, that is, a system of $n$ coupled oscillators, where an oscillator $j$ is connected with a unit coupling to its neighbors in $G$, and the natural frequency of each oscillator is zero. Let the state of the system be $\theta = (\theta_1, \ldots, \theta_n)^{T} $, where $T$ denotes transpose. Then, the dynamics of the system is governed by differential equations defined on $\mathbb{T}^n = \left( \mathbb{R}/ 2\pi \mathbb{Z} \right)^n$ of the form:
\begin{equation} \label{KMeqns}
\dot{\theta}_j = \sum_{ i \sim j} \sin( \theta_i -\theta_j), 
\text{ for } j \in [n],
\end{equation}
where $[n] = \{1, \ldots, n\}$, and the sum is taken over all neighbors $i$ of $j$ in $G$, i.e., $ i \sim j$ denotes an edge $ij \in E(G)$. Note that in a fixed moving frame, the synchronized state, $(0, \ldots, 0)$, is always a stable equilibrium of \eqref{KMeqns}. Since \eqref{KMeqns} is a gradient system with energy function given by $\mathscr{E}_G(\theta) = \sum_{i \sim j} (1 -\cos(\theta_i -\theta_j) )$, the only attracting sets \eqref{KMeqns}  may have are stable equilibria, i.e., sync or other stable phase-locked solutions, the latter equilibria we call patterns. Hence in order to find out whether sync is not globally stable we need to check if the system possesses patterns. 

It was previously shown in \citep{Taylor12} that sync is globally stable for \eqref{KMeqns} on dense graphs, and it is easy to see that sync is also globally stable  on trees. Even though we lack a complete picture between these two extremes, there are a few graph types where it was shown that sync is not globally stable. For example, a ring graph on $n$ vertices with $k$ edges to nearest neighbors has non-globally stable sync for sufficiently large $n$ and small $k$ as was shown in \citep{Wiley} where the question of the size of basin of attraction of sync was also addressed. This example along with the fact for trees shows that a necessary condition for a graph to have a pattern is the existence of a cycle in the graph. For 3-regular (cubic) graphs of order up to 30 it was shown in \citep{LeeBard}  that the fraction of graphs with non-globally stable sync increases with $n$.

{The main problem in answering whether or not sync is globally stable is that a specific graph structure guaranteeing this property has not yet been found.}  Moreover, it was conjectured in \citep{TaylorNP} that finding whether \eqref{KMeqns} has non-global sync is an NP-hard problem.

In the next sections we introduce a way to answer the question for a wide range of sparse connected graphs. We reformulate the question in terms of electrical circuits. Parallels between Kuramoto flow and a flow on electrical networks have been drawn before, see, e.g., \citep{DorflerBullo, Timme}. However, system \eqref{KMeqns} allows us to go further in the analysis of global stability of sync. As we show in the next sections, it is enough to consider only the graph Laplacian in the search of patterns.
This is possible due to the fact that oscillators are dynamically identical, since the natural frequencies are zero and the coupling is fixed at one; the only difference between them is inherited from the underlying graph. Throughout the paper, we use networks and graphs interchangeably, as well as vertices and nodes; other graph theoretical terminology is standard and can be found, e.g., in \citep{Godsil}.



\begin{figure}[!t]
\center{\includegraphics[width=0.95\linewidth]{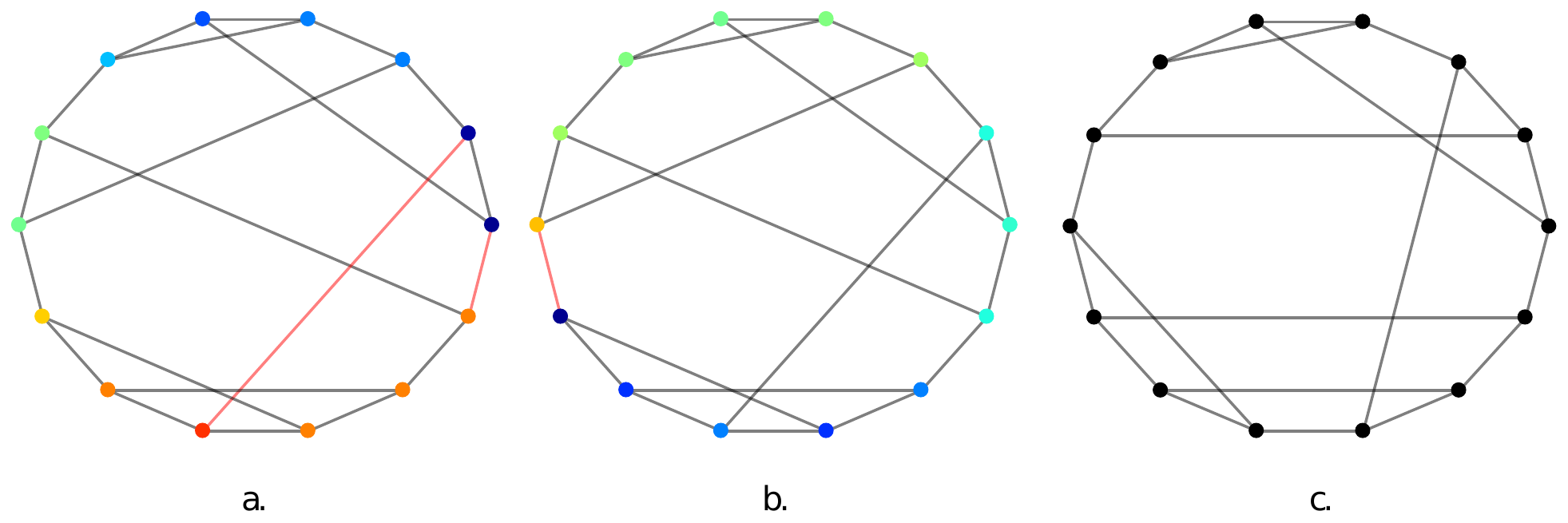}}
\caption{Graphs on a. and c. are cospectral cubic graphs. The Kuramoto model admits a pattern on graph a., i.e., sync is not globally stable, and does not have a pattern on graph c., i.e., sync is globally stable. The vertex-color-coded pattern on the graph which is obtained by using the red edges as boundary edges is shown on a. The colors of vertices on b. correspond to the initial condition (IC) obtained by using the red edge as a boundary edge. Starting from this IC the system converges to sync. (Color online.)}
\label{CoSpectral}
\end{figure}

\section{Spectra don't tell the full story}
It is known, see, e.g., \citep{Brouwer}, that the spectrum of the adjacency matrix of a graph does not define a graph uniquely, i.e., there are non-isomorphic graphs that share the same spectrum, which are called cospectral graphs. It is easy to see that if two $d$-regular graphs are cospectral then the eigenvalues of their Laplacians also coincide. In order to answer the question whether a given graph has a pattern, one might hope to rely on the spectrum of matrices associated with the graph. For example, the synchronizability of a network is defined by the ratio of the first and the last positive eigenvalues of the Laplacian in the master stability function \citep{Pecora}. 

However, the existence of phase-locked solutions, and hence the global stability of synchrony, 
{
most likely
}
cannot be addressed solely by considering the spectrum of the adjacency matrix or of the Laplacian, as the following example shows. Consider the graphs shown in Fig.~\ref{CoSpectral}a and Fig.~\ref{CoSpectral}c. These are two cubic graphs on 14 vertices that are cospectral; however, there is a pattern for the Kuramoto model on the graph shown on Fig.~\ref{CoSpectral}-a, 
{
but either the graph from Fig.~\ref{CoSpectral}-c has no patterns or, if it does, the total volume of the basin of attraction of said patterns is less than $10^{-4}$
}
(we used the dataset of cubic graphs of order at most 18 previously studied in \citep{LeeBard}, where stable equilibria were searched using Monte Carlo simulations). 
{Indeed, synchronizability is a local measure and existence of patterns is a global question.}

\begin{figure}[!t]
\center{\includegraphics[width=0.6\linewidth]{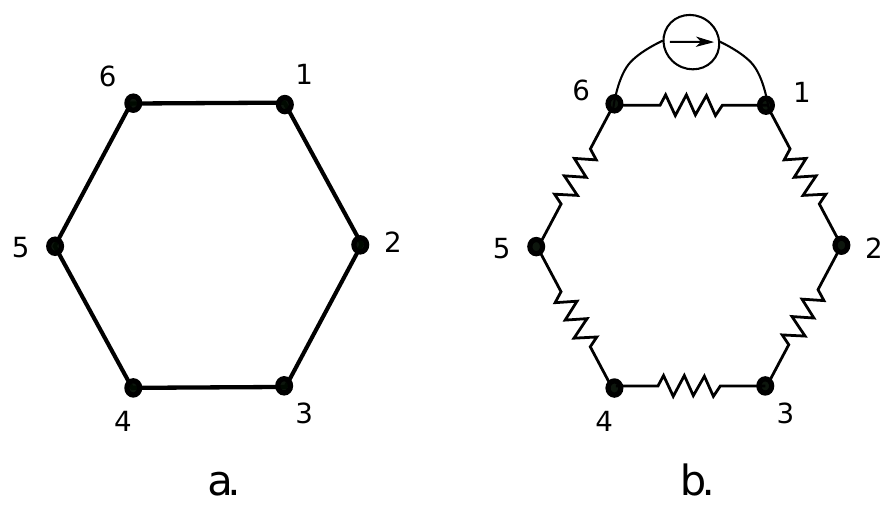}}
\caption{a. Ring graph (6-cycle); b. Electric circuit with current-source on 6-cycle graph where each edge is a unit resistor.}
\label{C6example}
\end{figure}

\section{Example}\label{ExampleSect}
Phase-locked solutions of the Kuramoto model are best understood on the $n$-cycle, i.e., the cycle graph with $n$ vertices, see, e.g., \citep{Wiley}. It is easy to check that $2\pi(0,\frac{1}{n}, \frac{2}{n}, \ldots, \frac{n-1}{n} )^{T}$ is a stable equilibrium of the Kuramoto model on the $n$-cycle for $n>4$. In fact, there is $Q = Q(n) < n$, such that for any $q\in\{ -Q,\ldots, Q \}$, the state $2\pi q(0,\frac{1}{n}, \frac{2}{n}, \ldots, \frac{n-1}{n} )^{T}$ is a stable equilibrium, also known as twisted state \citep{Wiley}, and $q$ is the winding number of the sate. 

In order to illustrate our central idea, we consider here the Kuramoto model on a ring graph with $n=6$ vertices shown in Fig.~\ref{C6example}a. We can find the pattern of this graph, if we think of the graph as a resistor network where each edge represents a resistor of 1~$\Omega$. Additionally, we add a current source of $i$ $A$ in parallel with an edge as shown on Fig.~\ref{C6example}b. We recall that in the Kuramoto model we consider, all natural frequencies are zero, i.e., there is not a constant term in the vector field that could be considered as an input. We repeat this because the resistor network has an input - the current source. What we have now is a linear electric circuit, and we can use the standard circuit analysis. Applying Kirchhoff Voltage Law and Kirchhoff Current Law to the circuit, one can find potentials $\phi_i$ at each vertex. Assuming that the current is $i=2 \pi~A$, we obtain voltages across each resistor, and setting the potential at the first branch point to $\phi_1 =0$, we obtain potentials that correspond to the phases in a pattern on 6-cycle.  In the remainder of this paper, we will use this idea to construct patterns on sparse graphs.


\section{Graph-theoretic preliminaries} \label{graphDef}
We can identify graphs by associated matrices. Consider a simple unweighted graph $G$ of order $n = |V(G)|$ and with size $m = |E(G)|$. Let $A = A_G$ be the adjacency matrix of $G$, i.e., symmetric square matrix with $A_{ij} =1$ if $ij\in E(G)$ and 0 otherwise. Then, the graph Laplacian of $G$ is defined as $L = L_G = D_G - A_G$, where $D_G$ is the diagonal matrix of degrees of vertices in $G$. (There should not be any confusion in what follows when we drop subindex $G$ in matrix notations.) If we randomly assign directions to all edges in a graph then we can express the Laplacian in a different form using the signed (edge-vertex) incidence matrix $B_{m \times n}$ defined as
\begin{equation}
B(e,v)  =
    \begin{cases}
        1, & \text{if $v$ is the head of $e$;} \\
        -1, & \text{if $v$ is the tail of $e$;} \\
	0, &  \text{otherwise.}
    \end{cases}
\end{equation}
Then the Laplacian is given by $L = B^{T} B$. We can generalize these definitions to weighted graphs. Let $W = \text{diag}(w_1, \ldots, w_m)$ be a diagonal matrix that contains all weights of a weighted graph $G_w$, then $L_w  = B^{T} W B$.

\section{Kirchhoff's laws for the Kuramoto model}
{
Let the Kuramoto model be defined on a graph with (signed) incidence matrix $B$. Then, we can rewrite the system \eqref{KMeqns} as in, e.g., \citep{Jadbabaie},
\begin{equation} \label{KMVector}
\dot{\theta} = B^T \sin( B \theta),
\end{equation}
where sine is taken element-wise. Then the equilibria of \eqref{KMVector} are solutions of $B^T \sin( B \theta) = 0$. If $\sin( B \theta) \neq 0$, that is, the corresponding solution is not the synchronized state, then $\sin( B \theta)$ is in the kernel of $B$, i.e., it lies in the cycle space of the graph. Using a basis of the cycle space $\{u_1, \ldots, u_{m-n+1}\}$ we can write
\begin{equation} \label{cyclSpac}
B \theta = \sin^{-1} \left[\sum_j c_j u_j \right],
\end{equation}
for some $c_j$, where we have omitted terms that are multiples of $\pi$ after taking $\sin^{-1}$.  If we left-multiply both sides of \eqref{cyclSpac} by $B^T$ then we get 
\begin{equation} \label{LaplcyclSpac}
L \theta =  B^T \sin^{-1} \left[\sum_j c_j u_j \right],
\end{equation}
where $L$ is the graph Laplacian. 

In section~\ref{ExampleSect}, we considered an example of a cycle graph and showed that a pattern can be obtained as a solution of an electric network problem. In general, an electric network problem and its solution, which relies on Kirchhoff's laws, can be written in terms of the Laplacian of the underlying graph of a circuit as
\begin{equation} \label{eq}
L \phi = i 
	\text{  and  } 
\phi = L^{+} i, 
\end{equation}
respectively, where $\phi = (\phi_1, \ldots, \phi_n)^{T}$ is the vector of potentials, $i$ is the applied current, and $L^{+}$ is the Moore-Penrose pseudo-inverse of $L$.

Based on a similarity between  \eqref{LaplcyclSpac} and \eqref{eq}, we state without a proof the following
\newtheorem{ansatz}{Ansatz}

\begin{ansatz}\label{Ansatz}
Let $\theta^*$ be a stable phase-locked solution of \eqref{KMeqns}. Then there exists a point $\tilde{\theta}$ in the basin of attraction of $\theta^*$, possibly $\theta^*$ itself, such that
\begin{equation} \label{LinEqForPatt}
L\tilde{\theta} = s,
\end{equation}
where $s \bot \mathbf{1}$ and $\mathbf{1} = (1,\ldots,1)^{T}$. Moreover, $s_i = 2\pi  c_i$ for $c_i \in \mathbb{Z}$, where $i \in [n]$. 
\end{ansatz}
If for an edge $ij \in E(G)$ we have $\text{sign}(s_i s_j)  < 0$ then it is called a boundary edge while vertices $i$ and $j$ called boundary vertices, and $s$ is 
a boundary vector.

Based on \eqref{cyclSpac}, \eqref{LaplcyclSpac}, and \eqref{LinEqForPatt},  Ansatz~\ref{Ansatz} says that there is an initial condition with the sum of phase differences along at least one cycle in the graph changing by an integer factor of $2\pi$ that leads to an equilibrium, i.e., all patterns have a twist, and a boundary edge is an edge from such a cycle.
Cycles with this property we call the core of a pattern. 
Such cycles, which carry cycle flows with a fixed winding number, were previously considered in \citep{Jacquod17, Delabays, Timme}, where, in particular, they were used for deriving bounds on the number of stable fixed points. To see that our and previously introduced methods complement each other, we make a comparison based on the laws for electric circuits. We can write equation that cycle flows have to satisfy 
\[
P B \theta^* = \xi,
\]
where $P$ is a $r \times m$ matrix with rows representing all $r$ cycles in $G$, and $\xi$ is a vector of winding numbers of the cycles. As before $B$ is the signed incidence matrix, however, the signs have to be selected in a way so that the cycle flow is well defined. This equation is equivalent to Kirchhoff's voltage law after we take it modulo $2 \pi$. While in our case, we have from Ohm's law 
\[
B \tilde{\theta}  = R \iota + \eta,
\]
where $\iota$ is $m$-dimensional current vector, $R$ is $m \times m$ diagonal matrix of resistances, for identical resistance case we consider here $R$ is the identity matrix, and $\eta$ is an $m$-dimensional vector of inputs. Multiplying the last equation by $B^T$ from the left and using Kirchhoff's current law for the first term on the right hand side, we get
\[
B^T B \tilde{\theta} = B^T \eta,
\]
which gives \eqref{LinEqForPatt} after using properties outlined in Section~\ref{graphDef}.

For a given pattern, a subset of edges taken from each cycle in the core constitutes a set of boundary edges, i.e., we can consider any edge of a cycle in the core as a boundary edge.
For simplicity, a set of boundary edges is called the boundary of a pattern, and the size of the boundary is the number of edges $ij \in E(G)$ with  $\text{sign}(s_i s_j)  < 0$ for $s$ as in \eqref{LinEqForPatt}.  A phase-locked solution has a simple core if it is determined by one boundary edge, i.e., by one cycle in a graph; and it has a complex core otherwise. So phase-locked solutions with a simple core have a boundary of size one, and phase-locked solutions with a complex core have a boundary of size bigger than one.

Let  ${\cal S}$ be a set of boundary vectors that determines all phase-locked solutions  of \eqref{KMeqns} on $G$. It may happen that $s^1 = q s^2$, where $s^1, s^2 \in {\cal S}$ and some $q \in [n]$. That is, two distinct patterns are determined by the same boundary edges. This is the case for $q$-twisted states on $n$-cycles studied in \citep{Wiley}. 

An example of a graph with a phase-locked solution that has a simple core is the $n$-cycle graph. In particular, the 6-cycle graph considered in section~\ref{ExampleSect}, see Fig.~\ref{C6example}a, has a simple core and edge, e.g., 1-6 is a boundary edge. Note, however, that every edge of the $n$-cycle graph is a boundary edge that defines the same pattern; in this case, the core is the graph itself.

Now, let us go back to the electric circuits analogy. In order to find patterns we need to add current sources of $2 \pi$ $A$ in parallel to boundary edges. If, however, the same core corresponds to a  $q$-twisted pattern with $q \ge 2$ then we need to place current sources of $2 \pi q$ $A$ in parallel to the boundary edges.

Similar to \eqref{eq} we can find a point in the basin of attraction of a phase-locked solution of \eqref{KMeqns} as the solution of \eqref{LinEqForPatt}, that is,
\begin{equation} 
\tilde{\theta} = L^{+} s,
\end{equation}
where $L^{+}$ is the Moore-Penrose pseudo-inverse of $L$, i.e., $ L^{+} = \sum_{i=2}^{n} \frac{1}{\lambda_i} u_i  u_i^{T}$ with $\lambda_i$ and $u_i $ being the positive eigenvalues and the corresponding eigenvectors of $L$, and some boundary vector $s$.

Using graph Laplacian it easy to show that an equilibria is stable, when phase differences between adjacent vertices are less than $\pi/2$, even though we do not impose this condition later in the paper. To see this, let $L_w$ be the Laplacian of a weighted graph with edge weights $w$. Then we can think about the Jacobian $J(\theta^*)$ of the Kuramoto model evaluated at an equilibrium $\theta^*$ as a negative multiple of the Laplacian of a weighted graph $J(\theta^*) = - L_w$ for some $w$. Even though it is easy to show using Gershgorin disc theorem that an equilibrium is stable under the assumption on phase differences, one can do the same by considering quadratic form associated with $L_w$. Indeed, under the assumption that phase differences between adjacent vertices are less than $\pi/2$, every edge of the weighted graph is positive, i.e., $w_{i,j} = \cos (\theta_i - \theta_j) > 0$ for $|\theta_i - \theta_j|<\pi/2$, and if we think about $J(\theta^*)$ as an operator, then for any $x \in  \mathbb{R}^n$ we have
\begin{eqnarray}
x^{T} J(\theta^*) x &=& - x^{T} L_w x = - x^{T} B^{T} W B  x = 
 -|| W^{\frac{1}{2}} B  x ||_2^2 
 \nonumber
\\
&=& - \sum_{i \sim j} w_{i,j} (x_i - x_j)^2 \leq 0,
\end{eqnarray}
which implies that $J(\theta^*)$ is negative semidefinite, and the number of zero eigenvalues of $J(\theta^*)$ equals the number of connected components of $G_w$.
}

\subsection{Motivation for definitions} 
Let us comment on the introduced definitions of boundary vectors (vertices and edges). The graph Laplacian $L(G)$ can be considered as a discrete version of the continuous Laplace operator $\Delta$, which is defined on graph $G$. The solution to the Laplace equation $\Delta u=0$ is a harmonic function, similarly a function that belongs to the kernel of $L$ is called a harmonic function. For this reason, if a graph does not have a core, i.e., there are no boundary edges (vertices) and $s = {\bf 0}$, then the solution of \eqref{LinEqForPatt} always defines sync, which is equivalent to saying sync is globally stable.

We call $s$  a boundary vector because of the analogy with the Dirichlet boundary-value problem. The Dirichlet boundary-value problem for a graph requires us to find $\theta$ which is a solution of \eqref{LinEqForPatt} for a given $s$. Since we consider only connected graphs, it is easy to see that the Dirichlet boundary-value problem has solutions of the form $\theta = \tilde{\theta} + \beta {\bf 1} $, where $\beta\in \mathbb{R}$ and ${\bf 1}$ is the vector of all ones. For a fixed labeling of vertices in $G$ we obtain a unique solution of \eqref{LinEqForPatt} by setting the phase of, say, the first vertex to zero, $\tilde{\theta}_{1} = 0$. Recall that phase-locked solutions are usually considered in a moving frame where the phase of one reference oscillator is pinned to zero.


\section{Where to look for patterns?} \label{WherePatterns}
A naive approach to search for patterns would be to consider every edge of each cycle in a graph as a boundary edge, i.e., consider vectors $B^T c^i$ as boundary edges, where $c^i$ is $m$-dimensional vector with $i$-th entry $2\pi$, then all linear combination of $B^T c^i$ and $B^T c^j$, etc. However, this is not efficient, even though we can find a unique solution of \eqref{LinEqForPatt} for any boundary vector $s$. How do we find $s$ for a graph, which is a boundary vector of a pattern if it exists? Here we provide a partial answer to this question. Once we know the full answer we can characterize all graphs that support a globally stable synchronized state of \eqref{KMeqns}.

\begin{figure}[!t]
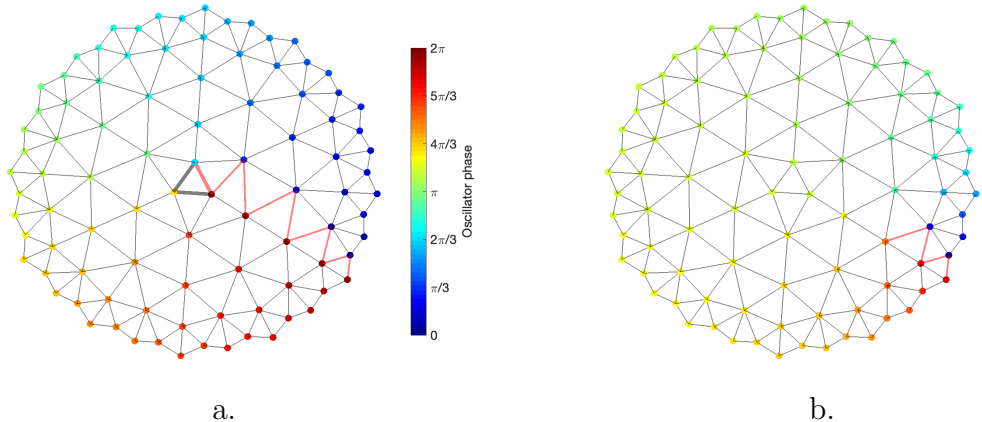

\begin{minipage}[h]{0.45\linewidth}
\center{\includegraphics[width=1.1\linewidth]{{{Figure_3a}.pdf}} \\ a.}
\end{minipage}
\hspace{1.6cm}
\begin{minipage}[h]{0.45\linewidth}
\center{\includegraphics[width=1.1\linewidth]{{{Figure_3b}.pdf}} \\ b.}
\end{minipage}
\caption{a. A pattern of a planar graph with a highly unstable subgraph; the phase difference between vertices of the inner triangle with bold edges is $2\pi/3$. The nine red edges are the boundary edges of the pattern, and the vertices are colored with respect to their phases in the pattern.  b. The vertices are colored with respect to initial condition (IC) obtained by considering the red edges as boundary edges. This is a wrong choice of boundary edges, the system converges to sync from this IC. (Color online.)}
\label{LongLinkPatt}
\end{figure}

Before we move to the description of some classes of graphs where it is easy to determine global stability of sync, we would like to note that the method described in this paper does not require that for a pattern all phase differences between adjacent vertices are less than $\pi/2$. The method works just fine when this condition is violated. Patterns that contain a pair of adjacent vertices with  phase differences greater than $\pi/2$ were called patterns with long links in \citep{LeeBard}. An example of a graph with pattern that contains a long link is shown on Fig.~\ref{LongLinkPatt}a. The phase difference between any pair of adjacent vertices of the inner triangle with bold edges is $2\pi/3$. This is, perhaps, the "longest" long link a pattern may have. This example also shows that the full graph can stabilize an unstable subgraph like 3-cycle with phases $(0,2\pi/3, 4\pi/3)$.

\begin{figure}[!t]
\center{\includegraphics[width=0.6\linewidth]{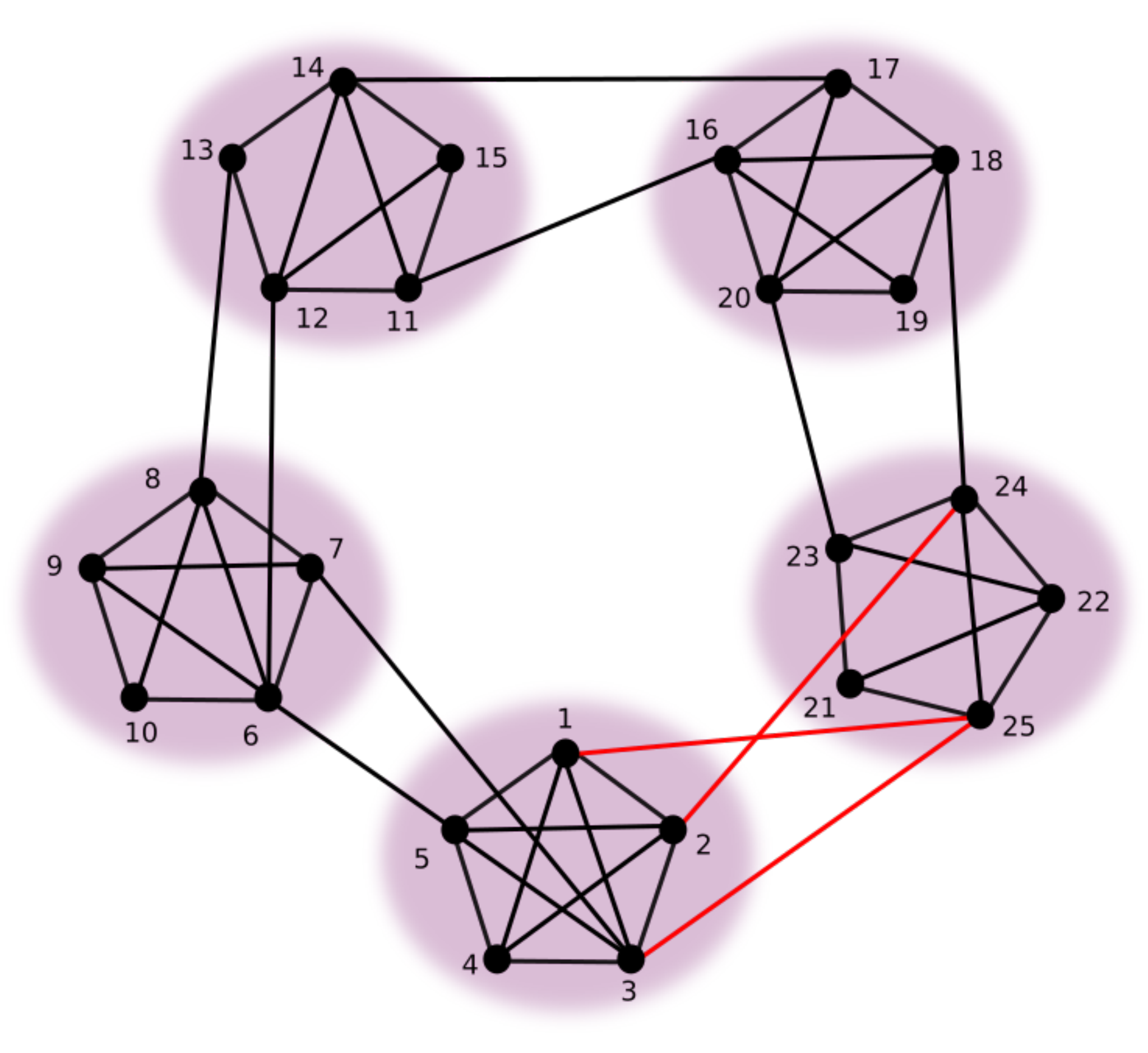}}
\caption{Cyclic graph clustering allows one to find boundary edges. For a sufficient number of clusters in cyclic graph clustering the boundary edges are the edges between any two adjacent clusters, e.g., the three red edges. For a given vertex labeling, the boundary vector of a pattern has the form $s = 2\pi(1,1,1,0,\ldots, 0,-1,-2)$. (Color online.)}
\label{ClustrPattern}
\end{figure}

The first class of graphs for which we can determine the boundary edges of a pattern consists of graphs that can be partitioned cyclically into clusters as shown in Fig.~\ref{ClustrPattern}, where vertices in a cluster can be connected to vertices from exactly two other clusters in the partition. We call such a partition a cyclic graph clustering. In this case, a community detection algorithm can be applied to determine the partition if it exists. Since the notion of a cluster (community), i.e., a densely connected subgraph, allows variability in the edge density inside a cluster, and as it turns out the number of clusters affects the existence of patterns on graphs with cyclic graph clustering, we have performed a computational analysis to identify the number of clusters in cyclic graph clustering and the densities of intra- and inter-cluster connectivities sufficient for the existence of patterns using stochastic block models, see Section 7. If the number of communities is big enough, which depends on the edge densities within and between the communities, then there is a pattern and the corresponding boundary edges are the edges between any two adjacent communities, e.g., the three red edges in Fig.~\ref{ClustrPattern}. Once we have determined the boundary edges we can write the corresponding  boundary vector by assigning positive values to vertices from one cluster and negative values to vertices from the adjacent cluster, which are the endpoints of the boundary edges.

If we consider the graph shown in Fig.~\ref{ClustrPattern} then the boundary vector which corresponds to select boundary edges is $s = 2\pi(1,1,1,0,\ldots, 0,-1,-2)$. The vertex labeled 25 gets $-2$ in its boundary vector component instead of $\pm 1$ as all the others since the number of incident boundary edges with the vertex is two and we choose negative signs for vertices in the cluster containing vertices 21-25 and positive signs for vertices in the cluster with vertices 1-5.

As we mentioned earlier, $L=B^TB$ for any orientations of edges in a graph. However, we would like to note that once we found cyclic graph clustering of a graph the orientations of edges in $B$ between any adjacent clusters have to be the same in order to write a boundary vector $s$ in the form $B^T p$, where $p$ is an $m$-dim $0,2\pi$ vector with nonzero entries corresponding to boundary edges. Otherwise, $p$ would not contain $\pm 2\pi$ to maintain a consistent network flow.

While it is more likely that graphs with large girths have patterns, even a graph with a large number of triangles and a small diameter may have non-globally stable sync as shown on Fig.~\ref{LongLinkPatt}a. The way the boundary edges were obtained for that graph is the same for all planar graphs, which is another type of graphs where we can determine the boundary edges of a pattern. Assume that we are given a planar graph with at least one cycle and a particular planar embedding. Then in order to determine whether sync is globally stable we need only to check boundaries that are determined as follows. All edges that intersect a line connecting the exterior region and a region bounded by a cycle are the candidate boundary edges. For example, the nine red edges on Fig.~\ref{LongLinkPatt}a are the boundary edges that determine the pattern. Therefore, it is enough to check at most $|E(G)|-|V(G)|+1$ boundary vectors. This upper bound is the number of cycles in a graph, or cyclotomic number, and since we consider only sparse graphs, this method would be more efficient than exploring the whole phase space. 
{
We also note here that example shown on Fig.~\ref{LongLinkPatt}a resembles stable rotating waves on a square grid studied in  \citep{PaulletErmentrout}. Since a square grid is also a planar graph we can use our algorithm to find a pattern on that graph. While the pattern Fig.~\ref{LongLinkPatt}a suggests that stable rotating waves can also be found on a 3-regular graph.
}

\begin{figure}[!t]
\center{\includegraphics[width=1\linewidth]{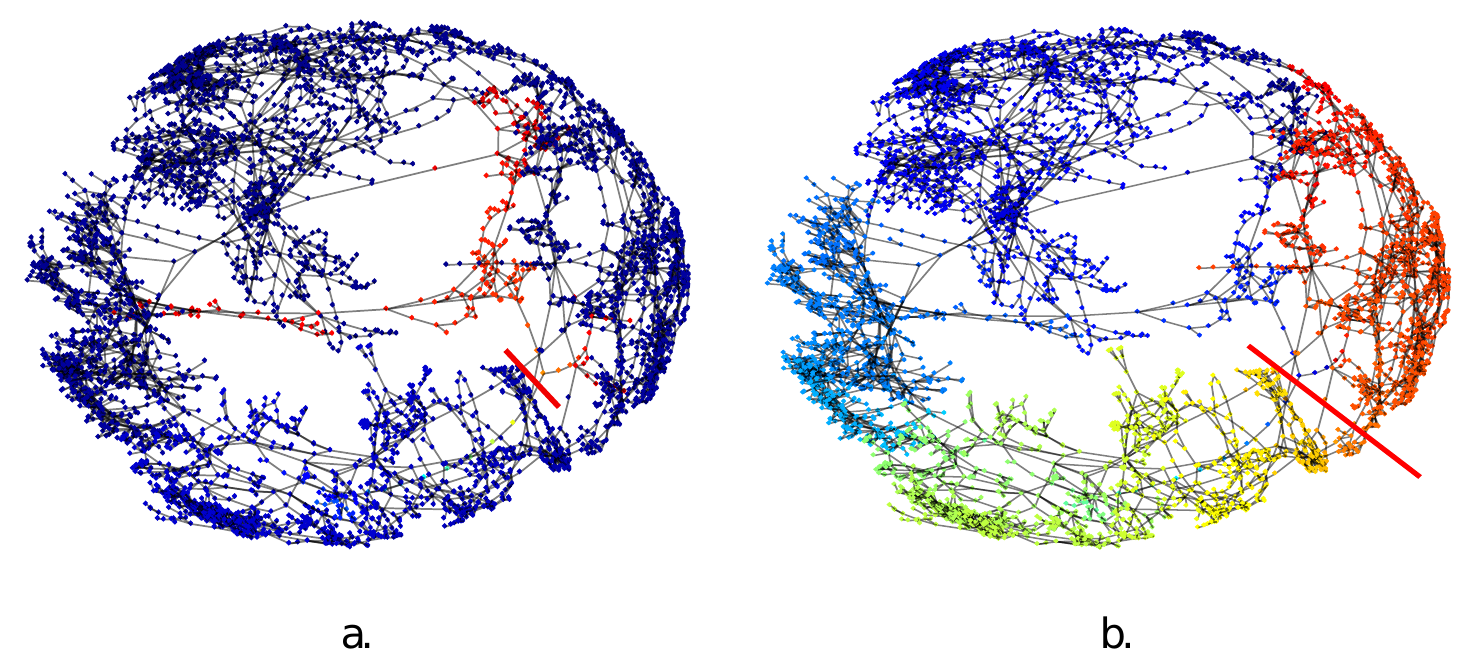}}
\caption{Small-world network on 4941 vertices with low density of triangles (power-grid network studied in \citep{WattsStrogatz}). Nodes are colored with respect to phases of nodes in two different patterns $\mod 2\pi$ (color is online) with $\theta_1$ fixed at zero. A pattern with a simple core is shown on a) and with a complex core on b). Boundary edges of a pattern are the edges of the graph which intersect the corresponding red line. (Color online.)}
\label{PowerGridGraph}
\end{figure}

In the case of large real networks with small number of triangles, like the power grid network shown on Fig.~\ref{PowerGridGraph}, we can determine the boundary edges via cyclic graph clustering. Note that we can relax the definition of a cluster by allowing a cluster to contain just one vertex in cyclic graph clustering. {In particular, in the presence of a sufficiently long path without the cut vertices of a supergraph containing the path}, we have that sync on such a graph is not globally stable as the path is part of a core of a pattern as in Fig.~\ref{PowerGridGraph}a.

However, if a wrong set of edges is used as a set of boundary edges then the Kuramoto model will converge to sync starting from the initial condition (IC) determined by the set. For example, Fig.~\ref{CoSpectral}b and Fig.~\ref{LongLinkPatt}b illustrate two examples of an IC starting from which the Kuramoto model converges to sync if the boundary edges are selected to be the red edges. The colors of vertices correspond to phases in the IC.

{
As a summary, we list below the algorithm for the identification of initial conditions on different types of graphs considered here.
\begin{itemize}
\item[] Input a graph $G$ with adjacency matrix $A$
\item[1.] If $G$ is planar do $2$, otherwise $3$
\item[2.] Fix a planar embedding of $G$
\item[2.1.] For a given inner-face, find a (minimum) number of edges needed to be deleted so that the inner-face and the outer-face become  1-connected.
\item[2.2.] The deleted edges are the boundary edges; go to step 4.
\item[3.] Find 2-connected subgraph of the graph 
\item[3.1.] If there exists an induced subgraph, which is a path graph of length at least 3, then an edge of the subgraph is a boundary edge; go to step 4.
\item[3.2.] Otherwise, apply community detection (clustering) algorithm, with a possible size of a cluster being as small as one vertex
\item[3.3.] Edges between any pair of clusters are the boundary edges;  go to step 4.
\item[4.] i.c. = $2*\pi$*MoorePenrose(A)*BoundVector
\end{itemize}
}
%

\section{Stochastic block models with cyclically ordered communities}
After recognizing the type of graphs, which may have patterns, we have performed the analysis of the Kuramoto model on stochastic block models with cyclically ordered communities to confirm the performance of the algorithm and to reveal the minimum number of clusters needed for the existence of patterns. 
{
Stochastic block model is a random graph, where the set of vertices is split into a (given) number of communities and for which the probability of an edge between a pair of vertices depends on the community labels of the vertices. The following rigorous definition of stochastic block model, along with a review of current developments, can be found in \citep{Abbe}. 
\newtheorem{definition}{Definition}
\begin{definition}
Let $n$ be a positive integer (the number of vertices), $k$ be a positive integer (the number of communities), $p = (p_1, \ldots, p_k)$ be a probability vector on $[k]$ (the prior on the k communities) and $W$ be a $k \times k$ symmetric matrix with entries in $[0, 1]$ (the connectivity probabilities). The pair $(X, G)$ is drawn under $\mathscr{G}(n, p, W)$ if $X$ is an $n-$dimensional random vector with i.i.d. components distributed under $p$, and $G$ is an $n-$vertex simple graph where vertices $i$ and $j$ are connected with probability $W_{X_i,X_j}$, independently of other pairs of vertices. 
\end{definition}
}
For our purpose, we consider stochastic block models where depending on the memberships of select vertices the probability of an edge is: $p_{w.c.}$ if the vertices belong to the same community, $p_{b.c.}$ if the vertices are from neighboring communities, where each community has exactly two neighboring communities, and zero otherwise. The outcome of such a stochastic block model is a graph with cyclic graph clustering.

\begin{figure}[!t]
\center{\includegraphics[width=1\linewidth]{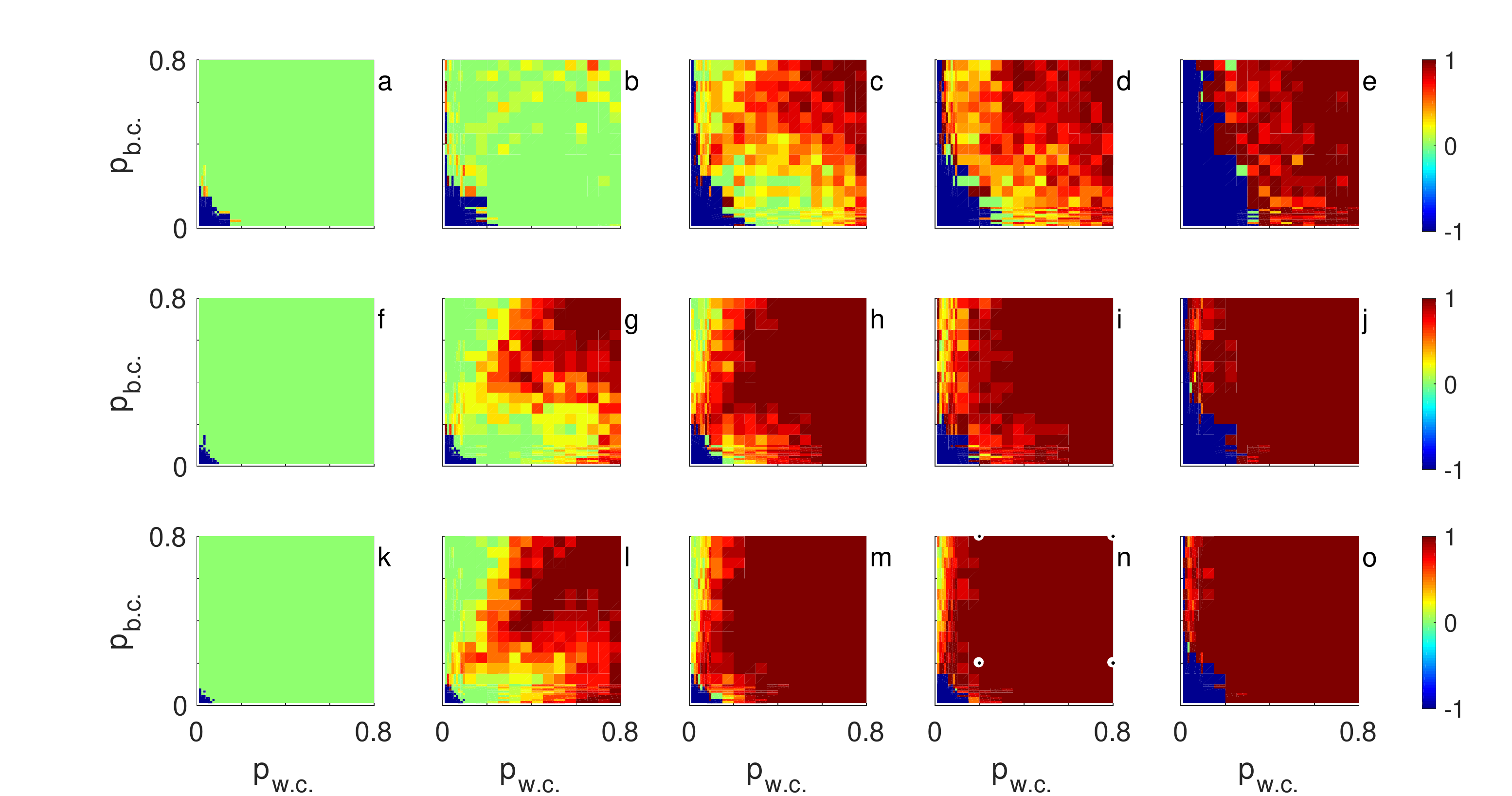}}
\caption{The average type of the final state in the Kuramoto model on stochastic block models under the condition on a generated graph being connected and with cyclic graph clustering. The types of a final state are: 0 if sync, 1 if a pattern, and -1 if a generated graph does not satisfy the condition (in this case we did not run the Kuramoto model). The edge probabilities within a community ($p_{w.c.}$) and between two consecutive communities in cyclic graph clustering ($p_{b.c.}$) vary 
from 0.01 to 0.1 with an increment 0.01, and from 0.1 to 0.8 with an increment 0.05. 
The graph order: in a-e is 50, in f-j is 100, and in k-o is 150. The number of communities in stochastic block models are 3, 4, 5, 6, and 10, which correspond to columns of subplots. For every combination of parameters we generated 10 graphs. Colors correspond to conditional averages, where the blue color means that all 10 graphs did not satisfy the condition. Four white circles in n define the values of parameters in Fig. 9. (Color online.) 
}
\label{50BlockModel}
\end{figure}

We consider stochastic block models of orders 50, 100, and 150, where the set of vertices is partitioned into clusters of sizes $3, 4, \ldots, 10$ with approximately equal number of vertices per cluster, where the probability of an edge within a cluster and between two consecutive clusters varies from 0.01 to 0.8 as described in the caption of Fig.~\ref{50BlockModel}. For each combination of parameters we generate ten realizations of graphs, and we run the Kuramoto model with an initial condition calculated as described in Section~\ref{WherePatterns} under the condition that a generated graph is connected and has cyclic graph clustering. 

The conditional averages of final states of the Kuramoto model on graphs with 3, 4, 5, 6, and 10 communities are shown on Fig.~\ref{50BlockModel} (see the caption for the description of a conditional average and other details). The blue region corresponds to parameters when the condition is not satisfied for all ten realizations. If there are only 3 clusters, the Kuramoto model mostly converges to sync from an identified initial condition, Fig.~\ref{50BlockModel} a, f, and k. However, as the number of clusters is increasing the Kuramoto model tends to have patterns, which are identified using the method described in Section~\ref{WherePatterns}, almost always when the underlying graph satisfies the condition, as shown in Fig.~\ref{50BlockModel} j, n, and o.

Based on simulations summarized on Fig.~\ref{50BlockModel}, even though there are a few instances when the Kuramoto model has a pattern on a graph with just three clusters and small probabilities of edges within and across clusters, Fig.~\ref{50BlockModel}~a, for large graphs ten clusters are mostly sufficient for the existence of patterns whenever the condition on connectedness and cyclic graph clustering is satisfied. 


\section{Destabilizing sync}
Based on remarks at the end of Section~\ref{WherePatterns}, i.e., if a graph has a sufficiently long induced subpath without cut vertices of the graph then the synchronized state is not globally stable as the cycle containing the path will be the (simple) core of a pattern. This suggests that we can destabilize the globally stable synchronized state by network changes like edge deletions. (We note that throughout this section, sync is locally stable, but will not be globally stable.  That is, when we refer to destabilizing sync, we mean it in the global sense.)  Among available edge-candidates we can choose the smallest number of edges to generate a sufficiently long induced subpath without cut vertices of the resulting graph.

Previously, a role of shortcuts on synchronization was discussed in the literature, see, e.g., \citep{WattsStrogatz}. However, the current paper shows that we do not need, in general, to delete a set of edges that necessarily would significantly increase  the diameter of a graph. Even local interventions that create a small but sufficiently long cycle in a graph may guarantee (global) destabilization of synchronized state by creation of a pattern.

\begin{figure}[!t]
\center{\includegraphics[width=0.5\linewidth]{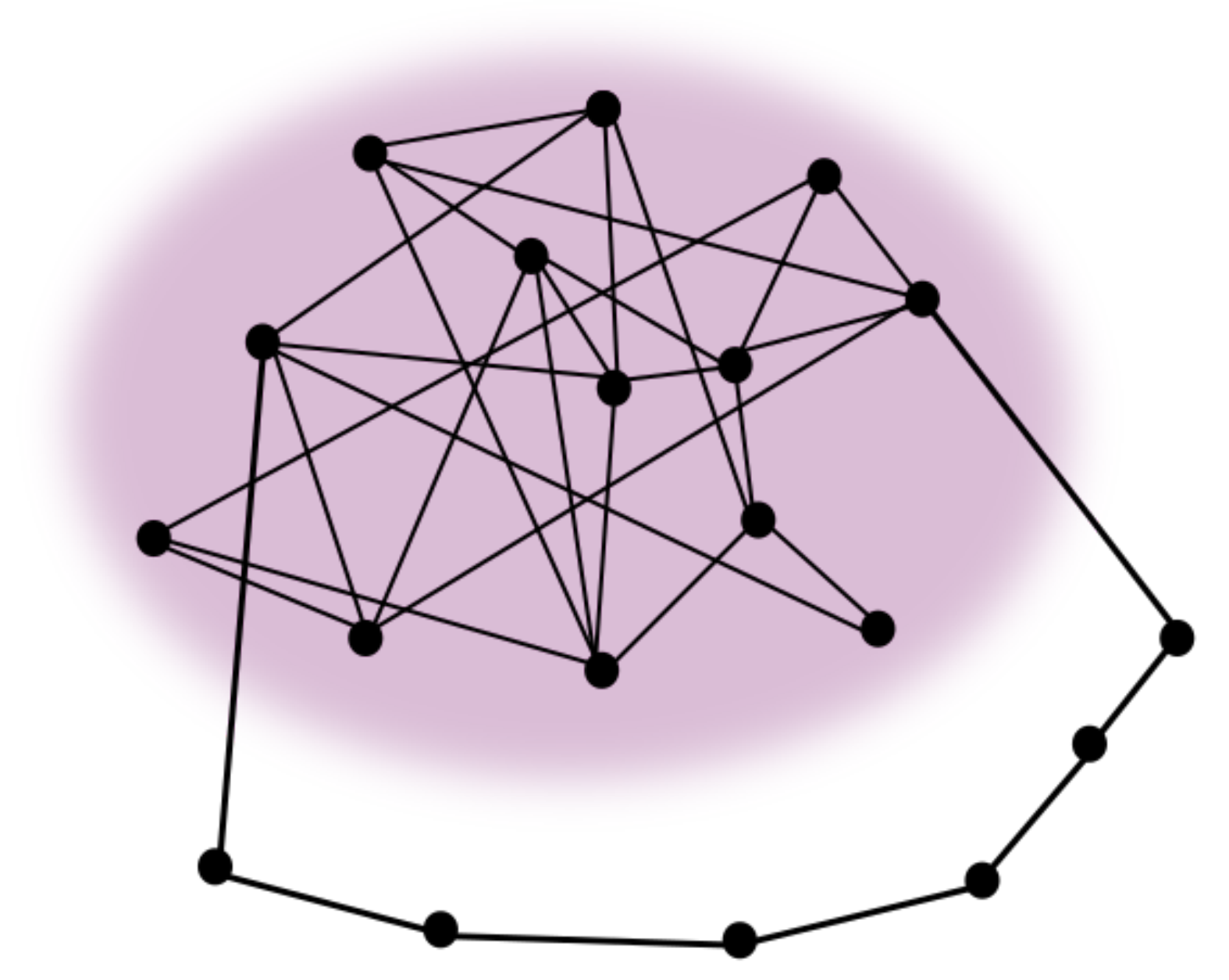}}
\caption{Appended path between two vertices of the graph shown in the shadowed region destabilizes sync and creates a pattern on the new graph.}
\label{destabSyncPic}
\end{figure}

Another way to destabilize the synchronized state is to append a sufficiently long path between a pair of vertices in a graph as shown in Fig.~\ref{destabSyncPic}. It is hard to compare directly the basin of attraction of sync in the original graph and in the modified one, since the new system has a higher dimension due to the increase of the number of vertices. Since we consider unweighted graphs and all natural frequencies are zero in \eqref{KMeqns} we can define dynamics for the new vertices similar to the others so that the new system is defined uniquely and is of the same type as the original one. While a shortcut itself should promote synchronization or at least not to decrease the basin of attraction, the subdivision of the shortcut, which leads to a small but sufficiently long path would destabilize the synchronized state. Since the subdivision of a shortcut, i.e., shortcut-path, would still, in general, decrease the diameter of a sparse graph, this illustrates that graph distance between vertices may have a different influence on synchronization. In contrast to subdivision, the edge contraction would not decrease synchronization and may lead to a graph with globally stable sync. 

\begin{figure}[!t]
\center{\includegraphics[width=0.6\linewidth]{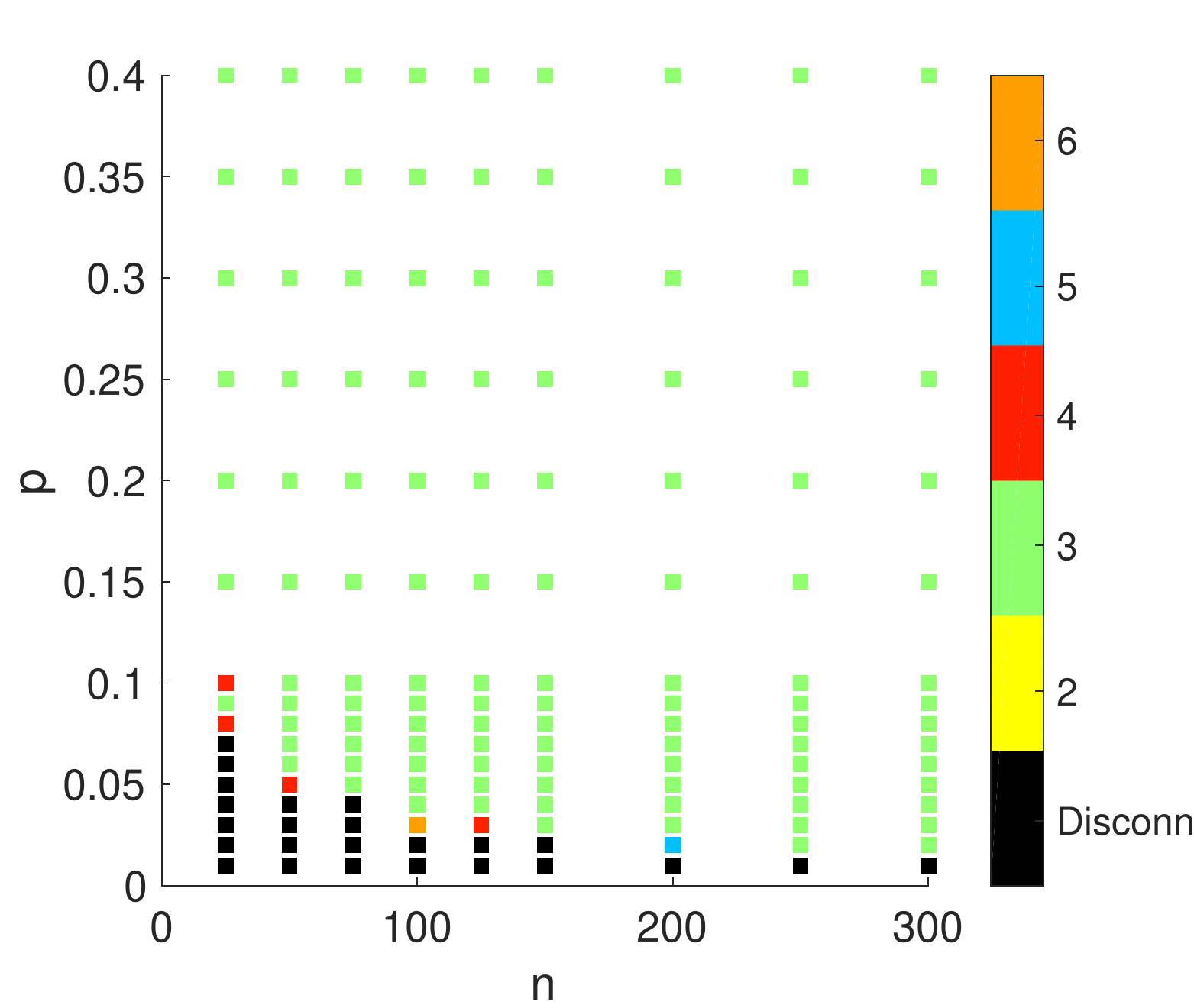}}
\caption{Sufficient length of augmented path added to two arbitrary vertices of an Erd\H{o}s-R\'enyi random graph $G_{n,p}$ needed to globally destabilize sync. { A sufficient length depicted by a color is defined as the minimum path lengths required to create a pattern on all ten realizations } of $G_{n,p}$ for a fixed combination of parameters $n$ and $p$, such that there is a pattern on all ten realizations of $G_{n,p}$ with longer augmented paths. (Color online.)}
\label{destabSyncGnp}
\end{figure}

To estimate how big an appended path should be to destabilize sync, we generate an Erd\H{o}s-R\'enyi random graph $G_{n,p}$ and augment a path of lengths $2, \ldots, 7$ to two arbitrary vertices of the graph. A sufficient length of a path appended to $G_{n,p}$ is shown in Fig.~\ref{destabSyncGnp}. When a resulting graph is connected then the sufficient length is usually three. The sufficient length is bigger near the boundary of the region where graphs become connected with the maximum being 6.


\section{Saddles come into play}
In Ansatz \ref{Ansatz} we stated that patterns can be found using boundary vectors with entries of the form $ 2\pi c_i$ where $c_i \in \mathbb{Z}$. What if $c_i \in \mathbb{R}$? First note that if $c_i=0$ for all $i$ then we will get sync as a solution of \eqref{LinEqForPatt}. Let ${c^*}$ be a vector with entries $c^*_i$ that corresponds to a pattern. It turns out that we can find a saddle that separates sync and the pattern using the boundary $s=2\pi \bar{c}$, where $\bar{c} = \alpha  c^*$ with $\alpha \in (0,1)$. In particular, if one considers graphs with simple cores like $n$-cycles then one will obtain 1-saddles, which were discussed in \citep{Lee1saddle}, in this way. 

Since the number of boundary vertices in $s$ is usually much less than the number of vertices in a graph, we fix column vectors of $L^+$ defined by those boundary vertices as seen from \eqref{LinEqForPatt}, and explore a lower dimensional subspace of the phase space by varying $c_i$. 

In the case of twisted patterns that share the same core one can also find saddles that separate the twisted patterns using this same idea. That is, if $s^{1} = q s^{2}$, where $q>1$ and $s^i = 2\pi \bar{c}^i$ for $ i \in \{ 1,2 \}$, are two boundary vectors. Then there is $\alpha \in (1,q)$ such that $\bar{c} = \alpha \bar{c}^{1}$ defines a saddle that separates the two twisted patterns with the same core.

\begin{figure}[!t]
\center{\includegraphics[width=0.95\linewidth]{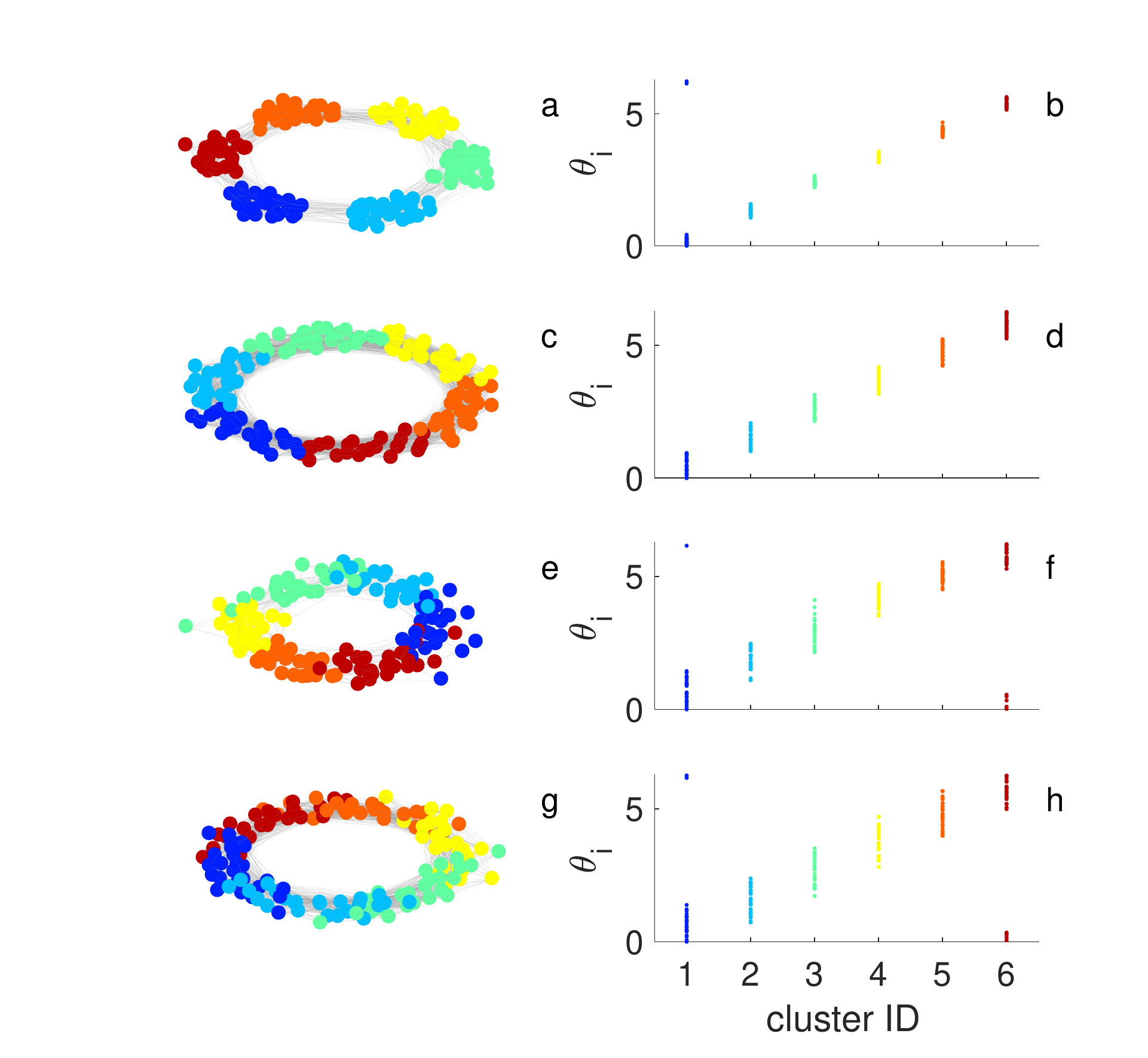}}
\caption{
{
Cluster synchronization as a function of $p_{w.c.}$ and $p_{b.c.}$. Four examples correspond to four combinations of $p_{w.c.}$ and $p_{b.c.}$ as shown by white circles on Fig.~\ref{50BlockModel}n, the other parameters are the same as those used for Fig.~\ref{50BlockModel}n. The first column contains an example of a graph generated with fixed set of parameters, vertices are colored according to their cluster membership. The second column shows the correlation between the phases of vertices in a pattern of the KM on the graph ($y$-axis) and their cluster memberships ($x$-axis), the color of a dot is the same as the color used for a cluster. Probabilities of edges are $p_{w.c.}=0.8$ and $p_{b.c.}=0.2$ for a and b,  $p_{w.c.}=p_{b.c.}=0.8$ for c and d, $p_{w.c.}=p_{b.c.}=0.2$ for e and f, and $p_{w.c.}=0.2$ and $p_{b.c.}=0.8$ for g and h. 
}
(Color online.)}
\label{ClustSync}
\end{figure}

\section{Conclusion}
In this paper we introduced a way to search for non-synchronous phase-locked solutions of the Kuramoto model on sparse graphs that takes advantage of the form of the system and provides a relation between structure of the underlying graph and stable equilibria in the phase space. This in turn helps to characterize graphs with globally unstable sync, in particular: graphs with sufficiently long induced subpaths, which do not contain cut vertices of supergraphs,  and graphs that permit cyclic graph clustering admit patterns.

In the way we find patterns, we first search for communities in the network and then use edges between adjacent communities. 
Two problems: community detection and the existence of patterns are related but not equal. As the example of the power grid network shows, Fig.~\ref{PowerGridGraph}b, a pattern can be used for separating the vertices of the network into communities based on similarity of phases in the pattern.
 However, the same network has a long induced subpath that defines a pattern as shown in Fig.~\ref{PowerGridGraph}a that purely separates vertices into communities. 
Another simple example, which illustrates that some patterns cannot be used for partitioning vertices into  communities based on phase similarity, is as follows. Consider two copies of the $n$-cycle graph with $n>5$ joined by a path. There is a pattern on this graph with vertices from the path having identical phases, and the phases of vertices on a cycle the same as in a pattern on the $n$-cycle. For a sufficiently long path the $n$-cycles are different communities in the graph. However, for each vertex of a cycle there is a vertex in the other cycle with the same phase, which would be considered as members of one community if the vertices were partitioned with respect to phase similarity.

In addition to the synchronization of the network, it was shown that one may have cluster synchronization, where vertices from the same cluster have identical phases \citep{Pecore-cluster-synchronization}. Early works suggest that this phenomenon may appear in networks with "high" symmetry \citep{Golubitsky-Stewart}. As we show here, if a graph allows a cyclic graph clustering with sufficient number of clusters 
 then sync is not globally stable and we can find a pattern using edges between adjacent clusters as boundary edges. { The resulting pattern would have the form where different vertices in a given cluster are nearly synchronized if the edge density inside the cluster is high as seen from Fig.~\ref{ClustSync} a-b. The difference in phases in a cluster will increase if the density of edges in the cluster decreases as shown on Fig.~\ref{ClustSync} e-h}. In other words, we may have nearly cluster synchronization, which depends on the densities of edges within and across clusters but not necessarily on symmetry.

\section{Acknowledgment}
The work of G.B.E. is supported in part by NSF~grant~$\#$~1712922.

%
%
%
%
%
%
%


%

\end{document}